\documentclass[12pt]{article}
\usepackage{amsfonts}

\usepackage{graphics}
\usepackage{indentfirst}
\usepackage{cite}
\usepackage{latexsym}
\usepackage{amsmath}
\usepackage{amssymb}
\usepackage[dvips]{epsfig}
\usepackage{amscd}

\usepackage{hyperref}
\usepackage{bibtopic}
\newtheorem {theorem}{Theorem}[section]
\newtheorem {lemma}{Lemma}[section]
\newtheorem {corollary}[theorem]{Corollary}

\newtheorem {remark}{Remark}[section]
\newtheorem {definition}{Definition}[section]

\usepackage{graphics}
\usepackage{indentfirst}
\usepackage{latexsym}
\usepackage{amsmath}
\usepackage{amssymb}
\usepackage[dvips]{epsfig}
\usepackage{amscd}
\usepackage{cite}

\def\bnn{\begin{eqnarray}}
\def\enn{\end{eqnarray}}
\def\bnnn{\begin{eqnarray*}}
\def\ennn{\end{eqnarray*}}
\def\la{\label}
\def\no{\nonumber\\}
\numberwithin{equation}{section}

\def\ba{\begin{aligned}}
\def\ea{\end{aligned}}
\def\bn{\begin{enumerate}}
\def\en{\end{enumerate}}
\def\be{\begin{equation}}
\def\ee{\end{equation}}

\def\lap{\triangle}

\def\g{\gamma}

\def\O{\Omega}

\def\ep{\varepsilon}

\def\p{\partial}

\def\norm[#1]#2{\|#2\|_{#1}}
\def\p{\partial}

\vspace{1cm}

\title{ Blowup   Criterion for the Compressible Flows with Vacuum States}

\author{  Xiangdi H{\small UANG}$^{a,c},$  Jing L{\small I}$^{b,c},$ Zhouping  X{\small IN}$^c$\thanks{
  Email addresses: xdhuang@ustc.edu.cn(X. Huang),\quad\quad\quad\quad    ajingli@gmail.com(J.
  Li),
zpxin@ims.cuhk.edu.hk(Z. Xin)}
 \\  {\normalsize  a. Department of Mathematics,}  \\  {\normalsize  University of Science and Technology of China,} \\
  {\normalsize  Hefei 230026,
 P. R. China}
 \\
   {\normalsize  b. Institute of Applied Mathematics, AMSS,} \\   {\normalsize Academia Sinica,
  Beijing 100190,
 P. R. China  }\\
  {\normalsize   c.  The Institute of Mathematical Sciences,} \\  {\normalsize  The Chinese University of Hong Kong, Hong
  Kong}}

\date{}

\begin{document}
\maketitle

\begin{abstract}

 We prove that the maximum norm of the deformation tensor of velocity gradients   controls
the possible breakdown of  smooth(strong)  solutions for the
3-dimensional compressible Navier-Stokes equations, which will
happen, for example, if  the initial density is compactly supported
\cite{X1}.  More precisely, if a solution of the compressible
Navier-Stokes equations is initially regular and loses its
regularity at some later time, then  the loss of regularity implies
the growth without bound of the deformation tensor as the critical
time approaches.  Our result is the same as Ponce's criterion for
3-dimensional incompressible Euler equations (\cite{po}).  Moreover,
our method can be generalized  to   the full Compressible
Navier-Stokes system which improve the previous results. In
addition, initial vacuum states are allowed in our cases.

\end{abstract}


\section{Introduction}
 The time evolution of the density and the velocity of a general
viscous compressible barotropic fluid occupying a domain
$\Omega\subset R^3$ is governed by the compressible Navier-Stokes
equations
\begin{equation}\label{a1}
\left\{ \ba
& \p_t\rho + {\rm div }(\rho u)=0,\\
& \p_t(\rho u) + {\rm div }(\rho u\otimes u) -\mu\lap u-(\mu +
\lambda)\nabla({\rm div }u) + \nabla P(\rho)=0 ,\ea \right.
\end{equation}
where $\rho, u, P$ denotes the density, velocity and pressure
respectively. The equation of state is given by \bnnn\la{a2}
 P(\rho) = a\rho^{\g}\quad (a>0,\g>1),
 \ennn
$\mu$ and $\lambda$ are the shear viscosity and the bulk viscosity
coefficients  respectively. They satisfy the following physical
restrictions: \be \la{a9}\mu>0, \lambda + \frac{2}{3}\mu\ge 0 .\ee
The equations (\ref{a1}) will be studied with  initial conditions:
\bnn (\rho,u)(x,0)=(\rho_0, u_0)(x),\la{a3} \enn and three types of
boundary conditions:

 1) Cauchy problem:   \bnn\la{b1}\O=R^3 \mbox{  and (in some weak
sense) $\rho,u$ vanish at infinity;}\enn

2) Dirichlet problem: in this case, $\O$ is a bounded smooth domain
in $R^3,$ and \bnn \la{b2} u=0 \mbox{ on }{\p\O};\enn

3) Navier-slip boundary condition: in this case, $\O$ is a bounded
smooth domain in $R^3,$ and \bnn\la{b3}u\cdot n=0,\quad {\rm
curl}u\times n=0\mbox{  on }   \p \O \enn
 where $n=(n_1,n_2,n_3)$ is the unit outward normal to $\p\O.$ The first condition in
$(\ref{b3}) $ is the non-penetration boundary condition, while the
second one  is also known in the form \bnn\la{b5}(\mathcal{D}(u)
\cdot n)_\tau=-\kappa_\tau u_\tau,\enn where $\mathcal{D} (u)$ is
the deformation tensor: \bnn\la{b7} \mathcal{D}(u)  =
\frac{1}{2}(\nabla u + \nabla u^t),\enn  and $\kappa_\tau$ is the
corresponding principal curvature of $\partial \O.$ Condition
(\ref{b5}) implies the tangential component of $\mathcal{D}(u) \cdot
n$ vanishes on flat portions of the boundary $\p \O.$ Note that
$\nabla u$ can be  decomposed as
  \bnn\la{z6} \nabla u=\mathcal{D}(u)+\mathcal{S}(u),\enn where
$\mathcal{D}(u)$ is the deformation tensor defined by (\ref{b7}) and
 \bnn\la{z5}\mathcal{S}(u)=\frac{1}{2}\left(\nabla u-\nabla u^t\right),
 \enn known as the rigid body rotation tensor. The tensors $\mathcal{D}(u)$ and
$\mathcal{S}(u)$ are respectively the symmetric and skew-symmetric
parts of $\nabla u.$

 There are huge literatures on
the large time existence and behavior of solutions to (\ref{a1}).
The one-dimensional problem has been studied extensively by many
people, see \cite{Kaz,Ser1,Ser2,Hof} and the references therein. The
multidimensional problem (\ref{a1}) was investigated by
Matsumura-Nishida \cite{M1}, who proved global existence of smooth
solutions for data close to a non-vacuum equilibrium, and later by
Hoff \cite{Hof,Hof2,Hof3}  for discontinuous initial data. For the
existence of solutions for arbitrary data,  the major breakthrough
is due to   Lions \cite{L1,L2}
   (see also Feireisl   \cite{F1}), where he obtains global
existence of weak solutions - defined as solutions with finite
energy - when the exponent $\gamma$ is suitably large. The main
restriction on initial data is that the initial energy is finite, so
that the density is allowed to vanish.

However, the regularity and uniqueness of such weak solutions
remains completely open. It should be noted that one cannot expect
too much regularity of Lions's weak solutions in general because of
the results of Xin(\cite{X1}), who showed that there is no global
smooth solution $(\rho,u)$ to Cauchy problem for (\ref{a1}) with a
nontrivial compactly supported initial density. Xin's  blowup result
(\cite{X1}) raises the questions of the mechanism of blowup and
structure of possible singularities: What kinds of singularities
will form in finite time?  What is the main mechanism for possible
breakdown of smooth solutions for the 3-D compressible equations?

We begin with the local existence of strong (or classical)
solutions. In the absence of vacuum, the local existence and
uniqueness  of classical solutions are  known  \cite{Na,se1}. In the
case where the initial density need not be positive and may vanish
in an open set, the existence and uniqueness of local strong (or
classical) solutions are proved recently in \cite{K1,K3,K4, K2, S2}.
Before   stating their local existence results, we first give the
definition of strong solutions.

\begin{definition}[Strong solutions]    $(\rho,u)$ is called a strong solution to (\ref{a1}) in $\O\times (0,T),$ if for some $q_0\in (3,6],$\be\la{a12} \ba
& 0\le \rho\in C([0,T ],W^{1,q_0}(\O)),\quad \rho_t\in C([0,T ],L^{q_0}(\O)),  \\
& u\in C([0,T ],D_0^1\cap D^2(\O))\cap L^2(0,T ;D^{2,q_0}(\O))\\
& u_t\in L^{\infty}(0,T_*;L^2(\O))\cap L^2(0,T ;D_0^1(\O)) ,\ea
 \ee and
$(\rho,u)$ satisfies (\ref{a1}) a.e.  in $\O\times (0,T).$
\end{definition}

Here and throughout this paper, we use the following notations for
the standard
 homogeneous and inhomogeneous Sobolev spaces.
   \bnnn \begin{cases} D^{k,r}(\O) =\{u\in
L^1_{loc}(\O):\norm[L^r(\O)]{\nabla^k u}<\infty\}, \\ W^{k,r}(\O) =
L^r(\O)\cap D^{k,r}(\O), \quad H^k(\O) = W^{k,2}(\O),\quad D^k(\O) =
D^{k,2}(\O),
\\  D_0^1(\O) = \left\{u\in L^6(\O):\norm[L^2(\O)]{\nabla u}<\infty, \mbox{ and }(\ref{b1})  \mbox{ or }(\ref{b2})
  \mbox{ or }(\ref{b3})  \mbox{ holds}\right\},
\\  H_0^1(\O) = L^2(\O)\cap D_0^1(\O),\quad \norm[D^{k,r}(\O)]{u} = \norm[L^r(\O)]{\nabla^k u}.\end{cases}\ennn

 In particular, Cho etc \cite{K1} proved the following result.
\begin{theorem}   If the initial data $\rho_0$ and $u_0$ satisfy \be\la{a10}
0\le\rho_0\in L^1(\O) \cap W^{1,\tilde{q}}(\O),\quad u_0\in
D_0^1\cap D^2(\O), \ee for some $\tilde{q}\in (3,\infty)$ and the
compatibility condition: \be \la{a11}-\mu\triangle u_0 - (\lambda +
\mu)\nabla{\rm div }u_0 + \nabla P(\rho_0) = \rho_0^{1/2}g \quad
\mbox{for  some } g\in L^2(\O) ,\ee then there exists a positive
time $T_1\in (0,\infty)$ and a unique strong solution $(\rho,u)$ to
the initial boundary value problem (\ref{a1})(\ref{a3}) together
with (\ref{b1}) or (\ref{b2}) or (\ref{b3})  in $\O\times (0,T_1].$
 Furthermore,  the
following blow-up criterion holds: if $T^*$ is the maximal time of
existence of the strong solution $(\rho,u)$ and $T^*<\infty$, then
\be \la{a13}\sup\limits_{t\rightarrow T^*}( \norm[H^1\cap
W^{1,q_0}]{\rho} + \norm[D_0^1]{u}) = \infty ,\ee with $q = \min
(6,\tilde{q}).$
\end{theorem}

 There are several  works (\cite{J1,J2,H1,X2,H4}) trying to establish   blow up criterions for the strong (smooth)
solutions to the   compressible Navier-Stokes equations. In
particular,  it is proved in \cite{J1} for two dimensions, if
$7\mu>9\lambda$, then\bnnn \la{a5}\lim_{T\rightarrow
T^*}\left(\sup_{0\le t\le T}\norm[L^{\infty}]{\rho}
+\int_0^T(\norm[W^{1,q_0}]{\rho} + \norm[L^2]{\nabla\rho}^4)dt
\right) = \infty ,\ennn where $T^*<\infty$ is the maximal time of
existence of a strong solution and $q_0>3$ is a constant.

Later, we \cite{H2,H4,X2} first establish a blowup criterion,
analogous to the Beal-Kato-Majda criterion \cite{B1} for the ideal
incompressible flows, for the strong and classical solutions to the
isentropic compressible flows in three-dimension: \be\la{a6}
\lim_{T\rightarrow T^*}\int_0^T\norm[L^{\infty}]{\nabla u}dt =
\infty, \ee under the stringent condition on viscous coefficients:
\be \la{a7}7\mu>\lambda. \ee

Recently,   the ideas of \cite{H2,H4,X2} has been generalized in
\cite{J2}  to establish a blowup criterion similar to $(\ref{a6}),$
under the same  assumption $(\ref{a7}),$ for the non-isentropic
fluids, that is, \be\la{a8} \lim_{T\rightarrow T^*}\left(
\sup\limits_{0\le t\le T}\|\theta\|_{L^\infty} +\int_0^T \norm
[L^{\infty} ]{\nabla u}dt\right) = \infty. \ee Very recently, in the
absence of vacuum, Huang-Li \cite{hl} succeeded  in removing the
crucial condition (\ref{a7}) of \cite{J2,H2,H4,X2} and established
blowup criterions (\ref{a6}) and \be\la{z8} \lim_{T\rightarrow
T^*}\int_0^T \left(\|\theta\|_{L^\infty}^2+\norm [L^{\infty}
]{\nabla u}\right)dt = \infty, \ee for isentropic and non-isentropic
compressible Navier-Stokes equations, under the physical
restrictions (\ref{a9}), respectively.

It should be noted that for ideal incompressible flows,
Beal-Kato-Majda \cite{B1} established a well-known blowup criterion
for the 3-Dimensional incompressible Euler equations that a solution
remains  smooth  if
 \bnn\int_0^T\norm[L^{\infty}]{\mathcal{S}(u)}dt\enn is bounded, where $\mathcal{S}(u)$ is  the rigid body rotation tensor defined by $(\ref{z5}).$ Later, Ponce
 \cite{po} rephrased  the Beal-Kato-Majda's theorem in terms of
 deformation tensor $\mathcal{D}(u),$ that is, the same results in \cite{B1}  hold
if \bnn\int_0^T\norm[L^{\infty}]{\mathcal{D}(u)}dt \enn remains
bounded. Moreover, as pointed out by Constantin\cite{C5},
 the solution is smooth if and only if $$\int_0^T\norm[L^{\infty}]{((\nabla u)\xi)\cdot \xi}$$ is bounded,
 where $\xi$ is the unit vector in the direction of vorticity ${\rm curl}u$. All these facts in \cite{B1,po,C5} show that the solution
 becomes smooth either the skew-symmetric
 or symmetric part of  $\nabla u$ is controlled.

Note that the results  in \cite{hl} are not so satisfactory in
two-fold:  one is that  the  results exclude initial vacuum states;
moreover, nothing is known from (\ref{a6}) about the natural
question: which part of $\nabla u,$ the symmetric part
$\mathcal{D}(u)$ or the skew-symmetric part $\mathcal{S}(u),$ will
become arbitrarily large as the critical time approaches?

The aim of this paper is  to  improve all the previous blowup
criterion  results for compressible Navier-Stokes equations by
removing the stringent condition (\ref{a7}), and allowing initial
vacuum states, and furthermore, instead of (\ref{a6}), describing
the blowup mechanism in terms of the deformation tensor
$\mathcal{D}(u).$ Our main result can be stated as follows:
\begin{theorem}\la{t1}
Let $(\rho,u)$ be a strong solution of the initial boundary value
problem (\ref{a1})(\ref{a3}) together with (\ref{b1}) or  (\ref{b2})
or (\ref{b3})  satisfying   $(\ref{a12}).$ Assume that the initial
data $(\rho_0,u_0)$ satisfies (\ref{a10}) and (\ref{a11}). If
$T^*<\infty$ is the maximal time of existence, then \be
\la{a14}\lim_{T\rightarrow
T^*}\int_0^T\norm[L^{\infty}(\O)]{\mathcal{D}(u) }dt = \infty, \ee
where $\mathcal{D}(u) $ is  the  deformation tensor defined by
(\ref{b7}).
\end{theorem}

A few remarks are in order:

\begin{remark} Theorem \ref{t1}  also holds for   classical solutions to
the compressible flows with initial vacuum, which  improves the
results of \cite{hl} to the case  where the initial density need not
be positive and may vanish in an open set. In addition, Theorem
\ref{t1} holds for all $\mu$ and $\lambda$   satisfying the physical
restrictions (\ref{a9}),  which removed the   condition (\ref{a7})
which is essential in the analysis in \cite{J2,H4,H2,X2}.
\end{remark}

\begin{remark}
In 1998, Xin\cite{X1} gave an life span estimate of classical
solutions to the compactly supported initial density of the Cauchy
problem  (\ref{a1})(\ref{a3})(\ref{b1}) at least in one dimension.
However, it's unclear which quantity  becomes infinite as the
critical time approaches. Theorem \ref{t1} shows that instabilities
can develop only if the size of the deformation tensor becomes
arbitrarily large.\end{remark}

\begin{remark} Theorem \ref{t1} gives a counter part of Ponce's result in \cite{po} for  the
incompressible flows.
  \end{remark}

Next,  we indicate that the  results in Theorem \ref{t1} can be
generalized    to the non-isentropic fluids described by
\begin{equation}\la{c0}
\left\{ \ba
& \p_t\rho + \text{div}(\rho u)=0,\\
& \p_t(\rho u) + \text{div}(\rho u\otimes u) -\mu\lap
u-(\mu + \lambda)\nabla(\text{div}u) + \nabla P=0,\\
& c_v[\p_t(\rho\theta) + \text{div}(\rho\theta u)] -
\kappa\triangle\theta + P\text{div}u = \frac{\mu}{2}|\nabla u +
\nabla u^T|^2 + \lambda(\text{div}u)^2,
 \ea \right.
\end{equation}
where $\theta$ is the absolute temperature, $P = R\rho\theta (R>0)$,
and   $\kappa\ge 0, R>0, c_v>0$ are physical constants.

The local existence of strong solutions with initial vacuum is
established in\cite{K3}, where it is essentially shown that for
initial data $(\rho_0,u_0,\theta_0)$ satisfying \be\la{c1} \ba &
0\le \rho_0\in W^{1,\tilde{q}}(\O)\quad
\text{for}\quad \text{some}\quad 3<\tilde{q}\le 6\\
& u_0\in H_0^1(\O)\cap H^2(\O), \theta_0\in H^2(\O),\theta_0\ge0,
\ea \ee and the compatibility conditions \be\la{c2} \ba &
\mu\triangle u_0 + (\mu+\lambda)\nabla\text{div}u_0 -
R\nabla(\rho_0\theta_0) =
\rho_0^{\frac{1}{2}}g_1,\\
& \kappa\triangle\theta_0 + \frac{\mu}{2}|\nabla u_0 + \nabla
u_0^t|^2 + \lambda(\text{div}u_0)^2 -R\rho_0\theta_0\text{div}u_0 =
\rho_0^{\frac{1}{2}}g_2, \ea \ee for some $g_1,g_2\in L^2(\O),$
furthermore, $\{x\in\O|\rho_0(x)=0\}$ being an open subset of $\O,$
there exist a $T_*>0$ and a unique strong solution $(\rho,u,\theta)$
on $[0,T_*]$ to the Cauchy  problem of (\ref{c0}), such that for any
$q_0\in (3,\tilde{q})$, \be\la{c3} \ba
& 0\le\rho\in C([0,T_*],W^{1,q_0}),\quad \rho_t\in C([0,T_*],L^{q_0}),  \\
& u\in C([0,T_*],D_0^1\cap D^2)\cap L^2(0,T_*;D^{2,q_0}),\\
& u_t\in L^{\infty}(0,T_*;L^2)\cap L^2(0,T_*;D_0^1),\\
& \theta\in C([0,T_*];H^2)\cap L^2(0,T_*;D^{2,q_0}),\theta>0,\\
& \theta_t\in L^{\infty}(0,T_*;L^2)\cap L^2(0,T_*;H^1). \ea
 \ee

By modifying the analysis for   Theorem $\ref{t1}$ and   in
\cite{hl},  one can  obtain  the following blowup criterion for the
full compressible Navier-Stokes system (\ref{c0}).
\begin{theorem}\la{t2}
Assume that the initial data satisfy $(\ref{c1})$ and $ (\ref{c2}).$
Let $(\rho,u,\theta)$ be a strong solution to the Cauchy problem of
$(\ref{c0}) $ satisfying   $(\ref{c3})$.
  If $T^*<\infty$ is the maximal time of existence, then
  \be\la{z9} \lim_{T\rightarrow T^*}\int_0^T
\left(\|\theta\|_{L^\infty}^2+\norm [L^{\infty} ]{
\mathcal{D}(u)}\right)dt = \infty. \ee
\end{theorem}

As aforementioned \cite{X1,Ch}, there are no global smooth solutions
for the compressible Navier-Stokes when the initial density is
compactly supported. We have the following corollary immediately.

\begin{corollary}
Assume that $(\rho_0,u_0,\theta_0)\in H^4(R^3)$  satisfy the initial
compatibility condition $(\ref{c2})$ such that   there exists a
finite number $0<r<\infty$ with $\text{supp}\rho_0\subset B_r$.
   Let $(\rho,u,\theta)$ be the corresponding classical solutions. Then there
  exists a time $T_*<\infty$, such that (\ref{z9}) holds.
\end{corollary}

We now comment on the analysis of this paper. Note that in all
previous works \cite{H2,H4,X2,hl}(see also \cite{J2}),  either the
  assumption $(\ref{a7})$ or the absence of vacuum played an important role in their analysis
  in order to obtain
an improved energy estimate which is essential not only for bounding
the $L^2$ norm of the convection term $F = \rho u_t + \rho
u\cdot\nabla u $ but also  for improving the regularity of the
solutions. Their method  depends on    the $L^\infty$-norm of
$\nabla u$ also. It is thus difficult to adapt their analysis here.
To proceed, some new ideas are needed. The key step in  proving
Theorem \ref{t1} is to derive the $L^2$-estimate on gradients of
both the density $\rho$ and the velocity $u.$ Observe that there are
two main difficulties here: one is due to the possible vacuum
states, the other is the strong nonlinearities of convection terms.
In order to overcome these difficulties, we will use the simple
observation that the momentum equations $(\ref{a1})_2$ become
``more" diffusive near vacuum if divided on both sides by $\rho$ as
long as $\rho$ remains bounded above which is guaranteed by the
boundedness of the temporal integral of the super-norm in space of
the deformation tensor. Thus a new energy estimate by using the
effective stress tensor will lead to a prior  estimates on the
$L^2$-norms of gradients of both the density and the velocity. A
simple combination of the above facts with the ideas used in
\cite{hl} then yields the blowup criterion for the full compressible
Navier-Stokes system (\ref{c0}).
 The details of the proof of   Theorems \ref{t1} and
 \ref{t2}  are given in Sections 2 and 3 respectively.

\section{Proof of Theorem \ref{t1}}

Let $(\rho,u)$ be a strong solution to the problem
 (\ref{a1})-(\ref{a9})  as described in Theorem \ref{t1}. First,   the standard
energy estimate yields \be\la{a16}\sup\limits_{0\le t\le
T}\left(\norm[L^2]{\rho^{1/2}u(t)}^2+\|\rho\|^\gamma_{L^\gamma}\right)
+ \int_0^T\norm[L^2]{\nabla u}^2dt \le C,\quad 0\le T<T^*. \ee
 To prove the theorem, we   assume otherwise that
\be\la{a15}\lim_{T\rightarrow
T^*}\int_0^T\norm[L^{\infty}(\O)]{\mathcal{D}(u) }dt\le C<\infty.
\ee Then (\ref{a15}), together with  $(\ref{a1})_1,$   immediately
yields the following $L^{\infty}$ bound  of the density $\rho.$
Indeed, on has
\begin{lemma}
  Assume that
\bnnn\la{a17}
  \int_0^T\norm[L^{\infty}]{div u}dt\le C, \quad 0\le T<T^*.
\ennn Then  \be\la{a18} \sup\limits_{0\le t\le T} \norm[L^{\infty}
]{\rho} \le C,\quad 0\le T<T^*. \ee
\end{lemma}
{\it Proof.} It follows from $(\ref{a1})_1$ that for $\forall p\ge
\gamma$, \be\la{a19}\p_t(\rho^p) + {\rm div }(\rho^pu) +
(p-1)\rho^p{\rm div }u = 0. \ee Integrating $(\ref{a19})$ over $\O$
leads to \bnnn\la{a}\p_t\int_{\O}\rho^pdx  \le
(p-1)\norm[L^{\infty}(\O)]{\text{div}u}\int_{\O}\rho^pdx, \ennn that
is, \bnnn\la{a20} \p_t\norm[L^p]{\rho} \le
\frac{p-1}{p}\norm[L^{\infty}(\O)]{\text{div} u}\norm[L^p]{\rho},
\ennn
 which implies immediately
  \bnnn\la{a21}
\norm[L^p]{\rho}(t)\le C ,\ennn with $C$ independent of $p$, so our
lemma follows.

   The   key estimates on $\nabla \rho$ and $\nabla u $ will be
given in the following lemma.
\begin{lemma}\la{z1}
  Under $(\ref{a15})$, it holds that for any $T<T^*$,
  \be\la{a30}
  \sup_{0\le t\le T}\left(\nabla\rho\|_{L^2}^2+\norm[L^2]{\nabla u}^2\right) + \int_0^T \norm[H^1]{\nabla u}^2 dt\le
  C.
  \ee

\end{lemma}

To prove Lemma \ref{z1}, we need the following lemma (see
\cite{bb}), which gives the estimate of $\nabla u$ by ${\rm div}u$
and ${\rm curl}u.$

\begin{lemma}\la{ppa1} Let $u\in H^s(\Omega)$ be a vector-valued function satisfying $u\cdot n|_{\partial\Omega}=0,$
where $n$ is the unit outer normal of $\partial\Omega.$ Then \bnn
\|u\|_{H^s}\le C(\|{\rm div}u\|_{H^{s-1}}+\|{\rm
curl}u\|_{H^{s-1}}+\| u\|_{H^{s-1}}),\enn for $s\ge 1$ and the
constant $C$ depends only on $s$ and $\Omega.$\end{lemma}

{\it Proof of Lemma \ref{z1}.} Multiplying $\rho^{-1}(\mu\triangle u
+ (\mu + \lambda)\nabla\text{div}u-\nabla P)$ on both sides of the
momentum equations $(\ref{a1})_2,$ integrating the resulting
equation over $\O,$ one has after integration by parts\bnn \la{a31}
&& \frac{d}{dt}\int_{\O }\frac{\mu}{2}|\nabla u|^2 +
\frac{\mu+\lambda}{2}(\text{div}u)^2dx +\int_{\O
}\rho^{-1}(\mu\triangle u + (\mu + \lambda)\nabla\text{div}u-\nabla
P)^2dx\no && \quad= -\mu\int_{\O }u\cdot\nabla u  \cdot\nabla\times
{\rm curl}u dx+(2\mu + \lambda)\int_{\O }u\cdot\nabla u \cdot
\nabla\text{div}u
 dx\no &&\quad\quad -\int_{\O }u\cdot\nabla u \cdot\nabla P dx - \int_{\O
}u_t\cdot\nabla Pdx ,  \enn due to $\Delta u=\nabla{\rm
div}u-\nabla\times{\rm curl}u.$ When  $u$ satisfies boundary
condition (\ref{b1}) or (\ref{b2}), we deduce from standard
$L^2$-theory of elliptic system that \bnn
\la{a56}\lefteqn{\|\nabla^2u\|^2_{L^2}-C\|\nabla
P\|_{L^2}^2}\no&&\le C\|\mu\triangle u + (\mu +
\lambda)\nabla\text{div}u\|^2_{L^2}-C\|\nabla P\|_{L^2}^2+C\|\nabla
u\|_{L^2}^2 \no&&\le C\|\mu\triangle u + (\mu +
\lambda)\nabla\text{div}u-\nabla P\|_{L^2}^2 +C\|\nabla
u\|_{L^2}^2\no&&\le C\int_{\O }\rho^{-1}(\mu\triangle u + (\mu +
\lambda)\nabla\text{div}u-\nabla P)^2dx+C\|\nabla u\|_{L^2}^2 ,\enn
due to $\rho^{-1}\ge C^{-1}>0.$  Lemma \ref{ppa1} yields that
(\ref{a56}) also holds for $u$ satisfying  boundary condition
(\ref{b3}) due to the following simple fact by (\ref{b3}): \bnnn
\|(2\mu+\lambda)\nabla{\rm div}u\|_{L^2}^2+\|\mu \nabla\times {\rm
curl}u\|_{L^2}^2=\|\mu\triangle u + (\mu +
\lambda)\nabla\text{div}u\|^2_{L^2}.\ennn

Next, we shall treat each term on the righthand side of (\ref{a31})
under the boundary condition (\ref{b3}), since the estimate  here is
more subtle than the other cases, (\ref{b1}) or (\ref{b2}), due to
the effect of boundary.

Using   (\ref{b3}) and the facts that $u\times {\rm
curl}u=\frac{1}{2}\nabla(|\nabla u|^2)-u\cdot\nabla u$ and
$\nabla\times (a\times b)=(b\cdot\nabla)a-(a\cdot\nabla)b+({\rm
div}b)a-({\rm div}a)b$, one gets  after integration by parts and
direct computations that \bnn\la{a32} \lefteqn{ \left|\int_{\O
}(u\cdot\nabla)u\cdot\nabla\times{\rm curl} udx\right|}\no &&
=\left|\int_{\O }{\rm curl}
u\cdot\nabla\times\left((u\cdot\nabla)u\right)dx \right|\no&&
=\left|\int_{\O }{\rm curl} u\cdot\nabla\times(u\times{\rm curl}u
)dx\right|\no&& =\left|\frac{1}{2}\int_\O|{\rm curl}u|^2{\rm
divu}dx-\int_\O{\rm curl}u\cdot\mathcal{D}(u)\cdot{\rm
curl}udx\right|\no& &\le C \|\nabla u\|_{L^2(\O)}^2
\|\mathcal{D}(u)\|_{L^\infty} , \enn and \bnn\la{a57}
\lefteqn{\left|\int_\O u\cdot\nabla u\cdot \nabla {\rm div}u
dx\right|}\no&&=\left|\int_{\p\O} u^i\p_i u^jn_j{\rm div}u
dS-\int_{\O}\nabla u:\nabla u^t\text{div}udx +
\frac{1}{2}\int_{\O}(\text{div}u)^3dx\right|\no&&\le\ep\|\nabla^2
u\|^2_{L^2 }+ C(\ep)\|\nabla u\|_{L^2 }^2\left(\|\nabla u\|_{L^2
}^2+\|\mathcal{D}(u)\|_{L^\infty}\right),\enn due to the following
simple fact: \bnnn \left|\int_{\p\O} u^i\p_i u^jn_j{\rm div}u
dS\right|&=&\left|\int_{\p\O} u^i\p_i (u\cdot n){\rm div}u
dS-\int_{\p\O} u^i u^j\p_in_j{\rm div}u dS\right|\no&=&
\left|\int_{\p\O} u^i u^j\p_in_j{\rm div}u dS\right|\no&\le&
C\|u\|^2_{L^4(\p\O)}\|{\rm div}u\|_{L^2(\p\O)}\no&\le& C\|\nabla
u\|_{L^2(\O)}^2\|\nabla  u\|_{H^1(\O)}\no&\le& C(\ep)\|\nabla
u\|_{L^2(\O)}^4+\ep\|\nabla^2 u\|^2_{L^2(\O)}+C(\ep),\ennn where
(\ref{b3}) and the Poincar\'{e} type inequality and the Ehrling
inequality have been used. Similarly,   \bnn
\la{a34}\lefteqn{\left|\int_{\O }u\cdot\nabla u\cdot\nabla
Pdx\right|}\no& &=\left|\int_{\p\O}u^i\p_iu^jn_j
PdS-\int_{\O}\p_ju^i\p_iu^jPdx-\int_{\O} u^i\p_i{\rm
div}uPdx\right|\no& &=\left|\int_{\p\O}u^iu^j\p_in_j
PdS+\int_{\O}\p_ju^i\p_iu^jPdx - \int_{\O }(\text{div}u)^2Pdx -
\int_{\O }u\cdot\nabla P\text{div}udx\right|\no&&\le C\|\nabla
u\|_{L^2}^2+\left| \int_{\O }u\cdot\nabla
P\text{div}udx\right|\no&&\le C\|\nabla u\|_{L^2}^2+
\norm[L^6]{u}\norm[L^3]{\text{div} u}\norm[L^2]{\nabla P}\no&&\le
C\|\nabla u\|_{L^2}^2+C\norm[L^2]{\nabla
u}^{\frac{5}{3}}\norm[L^{\infty}]{\mathcal{D}(u)
 }^{\frac{1}{3}}\|\nabla \rho\|_{L^2} \no&& \le C
\|\nabla\rho\|_{L^2}^2 \|\nabla u\|_{L^2}^2 +C\norm[L^2]{\nabla
u}^{2}\left( \norm[L^{\infty}]{\mathcal{D}(u)  }+1\right)+C,\enn
which yields also that \bnn\la{a36}\lefteqn{- \int_{\O
}u_t\cdot\nabla Pdx}\no& & =
 \frac{d}{dt}\int_{\O }P\text{div}udx - \int_{\O
}P_t\text{div}udx\no& & =   \frac{d}{dt}\int_{\O }P\text{div}udx +
\int_{\O }u\cdot\nabla P\text{div}udx + (\g-1)\int_{\O
}P(\text{div}u)^2dx \no& & \le \frac{d}{dt}\int_{\O }P\text{div}udx
+C\norm[L^2]{\nabla u}^{2}\left( \norm[L^{\infty}]{\mathcal{D}(u)
}+1\right)\no&&\quad+ C \|\nabla\rho\|_{L^2}^2 \|\nabla
u\|_{L^2}^2+C.\quad\enn Substituting (\ref{a56})-(\ref{a36})  into
$(\ref{a31}) $ gives that for  $\ep$ suitably small, \bnn\la{a37}
\lefteqn{\frac{d}{dt}\int_{\O }\left(\frac{\mu}{2}|\nabla u|^2 +
\frac{\mu+\lambda}{2}(\text{div}u)^2 - P\text{div}u\right)dx
 + C_0 \|\nabla^2 u\|_{L^2}^2 }\no&&\le
C \left(\|\nabla\rho\|_{L^2}^2+\|\nabla
u\|_{L^2}^2\right)\left(\norm[L^2]{\nabla
u}^{2}+\norm[L^{\infty}]{\mathcal{D}(u)}+1\right)+C.\quad\quad \enn
It remains to bound the $L^2$-norm of $\nabla \rho.$ To this end,
one can differentiate $(\ref{a1})_1$ and  then multiply the
resulting equation by $2\nabla\rho $ to get \bnn\la{a23}
 \lefteqn{\p_t|\nabla\rho|^2 + \text{div}(|\nabla\rho|^2u) +
 |\nabla\rho|^2\text{div}u}\no&& = -2(\nabla\rho)^t\nabla u\nabla\rho -
 2\rho\nabla\rho\cdot\nabla\text{div}u\no&& = -2(\nabla\rho)^t\mathcal{D}(u) \nabla\rho -
 2\rho\nabla\rho\cdot\nabla\text{div}u. \enn
Integrating  $(\ref{a23})$ over $\O $  yields \bnn
\la{a24}\p_t\left(\|\nabla \rho\|_{L^2}^2\right)\le
C\|\nabla\rho\|_{L^2}^2\|\mathcal{D}(u) \|_{L^\infty}+\ep\|\nabla^2
u\|_{L^2}^2+C(\ep)\|\nabla\rho\|_{L^2}^2.\enn Adding (\ref{a24}) to
(\ref{a37}), we deduce, after choosing $\ep$ suitably small and
using Gronwall's inequality, that  (\ref{a30}) holds. The proof of
Lemma \ref{z1} is completed.

Next step is   to improve the regularity of $\rho$ and $u$. We start
with some bounds on derivatives of $u$ based on above estimates.

\begin{lemma}\la{z2}
 Under the condition $(\ref{a15})$, it holds that
  \be\la{a42}
  \sup\limits_{0\le t\le T}\left(\norm[L^2]{\rho^{1/2}u_t(t)} +\norm[H^1]{\nabla u} \right) + \int_0^T\|\nabla u_t\|_{L^2}^2dt \le C,
  \quad 0\le T<T^*.
  \ee

\end{lemma}

{\it Proof.} Differentiating the momentum equations  $(\ref{a1})_2$
with respect to  $t$ yields \be \rho u_{tt} + \rho u\cdot\nabla u_t
- \mu\triangle u_t - (\mu + \lambda)\nabla{\rm div }u_t  = -\nabla
P_t -\rho_t u_t  - \rho u_t\cdot\nabla u-\rho_t u\cdot\nabla u . \ee
 Taking the inner product of the above equation with $u_t$ in $L^2(\O)$ and integrating by parts, one gets
\bnn \la{a43}\lefteqn{ \frac{1}{2}\frac{d}{dt}\int_{\O}\rho u_t^2dx
+ \int_{\O} \left(\mu|\nabla u_t|^2 + (\lambda + \mu)({\rm div
}u_t)^2\right)dx} \no& & = \int_{\O}P_t{\rm div }u_tdx-\int_\O
\rho(u_t\cdot\nabla u)\cdot u_t dx -\int_{\O} \rho
u\cdot\nabla\left(|u_t|^2 + u\cdot\nabla u\cdot u_t\right)dx
\no&&\le C\int_\O\left(|u||\nabla \rho|+|\nabla u|\right)|\nabla
u_t|dx+C\int_\O \left( \rho|u_t|^2|\nabla u| + \rho|u||u_t||\nabla
u_t|\right)dx\no&&\quad+ C\int_\O\left(|u||u_t||\nabla u|^2+
|u|^2|u_t||\nabla^2u| + |u|^2||\nabla u||\nabla
u_t|\right)dx\no&&=\sum_{i=1}^3I_i. \enn

Noticing that by (\ref{a30}) and Sobolev's inequality, one
has\bnn\la{a44} I_1&\le&{C\left(\|u\|_{L^\infty}\|\nabla\rho\|_{L^2}
+\|\nabla u\|_{L^2}\right)\|\nabla u_t\|_{L^2}} \no&\le&
C\norm[L^2]{\nabla u_t}\norm[H^1]{\nabla u}\no&  \le&
\ep\norm[L^2]{\nabla u_t}^2 + C(\ep)\norm[H^1]{\nabla u}^2.\enn
Similarly,  \bnn \la{a45}I_2&\le&{ C\norm[L^2]{\rho^{1/2} u_t
}\norm[L^6]{u_t}\|\nabla
u\|_{L^3}+C\norm[L^\infty]{u}\norm[L^2]{\rho^{1/2}u_t}\norm[L^2]{\nabla
u_t}} \no&  \le& C\norm[L^2]{\rho^{1/2} u_t }\norm[L^2]{\nabla
u_t}\|\nabla u\|_{H^1}\no& \le& \ep\norm[L^2]{\nabla u_t}^2 +
C(\ep)\norm[L^2]{\rho^{1/2}u_t}^2\|\nabla
u\|_{H^1}^2,\quad\quad\quad\quad \enn and \bnn \la{a46} I_3&\le&
C\norm[L^6]{u}\norm[L^6]{u_t}\norm[L^3]{\nabla u}^2+
C\norm[L^3]{u^2}\norm[L^6]{u_t}\norm[L^2]{\nabla^2 u}\no&&+
C\norm[L^2]{\nabla u_t}\norm[L^6]{\nabla u}\norm[L^3]{u^2} \no &\le&
C\norm[L^2]{\nabla u_t}\left(\norm[L^2]{\nabla u}\norm[L^6]{\nabla
u}+\|\nabla^2u\|_{L^2}\right)\no&\le& \ep\norm[L^2]{\nabla u_t}^2 +
C(\ep)\norm[H^1]{\nabla u}^2. \enn    We conclude from
(\ref{a43})-(\ref{a46}) that \bnnn
\lefteqn{\frac{1}{2}\frac{d}{dt}\int_{\O} \rho u_t^2 dx +\mu
\int_{\O}|\nabla u_t|^2dx}\no& & \le 6 \ep\norm[L^2]{\nabla u_t}^2 +
C(\ep) \|\nabla u\|_{H^1}^2 +
C(\ep)\norm[L^2]{\rho^{1/2}u_t}^2\|\nabla u\|_{H^1}^2 ,   \ennn
which, together with Gronwall's inequality,  implies that for $\ep$
suitably small, \be\la{a47} \sup\limits_{0\le t\le
T}\norm[L^2]{\rho^{1/2}u_t(t)}^2 + \int_0^T\|\nabla u_t\|_{L^2}^2 dt
\le C,
  \quad 0\le T<T^*,
\ee  due to the fact that   \bnnn \rho_0(x)^{\frac{1}{2}}u_t(x,t=0)
=
 \rho_0^{\frac{1}{2}}u_0\cdot\nabla u_0(x) - \rho_0^{\frac{1}{2}}g \in
 L^2(\O),
\ennn which comes from the compatibility condition (\ref{a11}).
  Moreover,
since $u$  satisfies \bnnn \begin{cases}\mu \Delta
u+(\mu+\lambda)\nabla{\rm div}u = \rho u_t + \rho u\cdot\nabla u +
\nabla P,\\ (\ref{b1}) \mbox{ or } (\ref{b2}) \mbox{ or } (\ref{b3})
   \mbox{ holds, } \end{cases} \ennn   similar to (\ref{a56}), one has  \bnnn\norm[L^2]{\nabla^2 u}&\le&
C(\norm[L^2]{\rho^{\frac{1}{2}}u_t} + \norm[L^\infty]{
u}\norm[L^2]{\nabla u} + \norm[L^2]{\nabla
P})\no&\le&C+C\|\nabla^2u\|_{L^2}^{1/2}
 .\ennn Hence, \bnn\la{z10} \sup\limits_{0\le T< T^*}\norm[H^1]{\nabla u} \le C.
\enn  Thus, Lemma $\ref{z2}$ follows from (\ref{a47}) and
(\ref{z10}) immediately.

Finally, the following lemma gives bounds of the first  derivatives
of the density $\rho$ and the second derivatives of the velocity
$u$.

\begin{lemma}\la{z3}
Under the condition $(\ref{a15})$, it holds that for  any $ q\in(3,6] $
 \bnn\la{a48}\sup\limits_{0\le t\le
T}  \norm[W^{1,q}]{\rho} \le C,\quad 0\le T<T^*.\enn
 \end{lemma}

{\it Proof.} In fact, $(\ref{a1})_1$ gives \bnnn \ba
& (|\nabla\rho|^q)_t + \text{div}(|\nabla\rho|^qu)+ (q-1)|\nabla\rho|^q\text{div}u  \\
 &+ q|\nabla\rho|^{q-2}(\nabla\rho)^t \mathcal{D}(u) (\nabla\rho) +
q\rho|\nabla\rho|^{q-2}\nabla\rho\cdot\nabla\text{div}u = 0 ,\ea
\ennn which yields for the case that (\ref{b1})   or (\ref{b2})
holds, \be\la{a50} \frac{d}{dt}\|\nabla\rho\|_{L^q} \le
 C(\norm[L^{\infty}]{\mathcal{D}(u) }+1)\norm[L^q]{\nabla\rho}  +
C\norm[L^q]{\nabla\text{div}u},  \ee and for the case that
 (\ref{b3}) holds, \be\la{a80} \frac{d}{dt}\|\nabla\rho\|_{L^q} \le
 C(\norm[L^{\infty}]{\mathcal{D}(u) }+\norm[L^q]{\nabla G}+1)\norm[L^q]{\nabla\rho}
 ,  \ee  with $G\triangleq (2\mu+\lambda)\text{div}u-P.$ Using the $L^p$-estimate of elliptic system, we have for the case that (\ref{b1})   or (\ref{b2})
holds,  \bnn\la{a49}
  \norm[L^q]{\nabla^2u}   &\le& C(\norm[L^q]{\rho u_t} + \norm[L^q]{u\cdot\nabla u} + \norm[L^q]{\nabla
 P})\no&\le&  C\left(\norm[L^2]{\sqrt{\rho}
u_t}^{(6-q)/(2q)}\|u_t\|_{L^6}^{(3q-6)/(2q)} +
\norm[L^\infty]{u}\norm[L^q]{\nabla u}  +
\norm[L^q]{\nabla\rho}\right)\no&  \le& C(\norm[L^2]{\nabla u_t}  +
\norm[L^q]{\nabla\rho}+1) ,  \enn  due to (\ref{a42}). When the
boundary condition (\ref{b3}) holds,  noticing that  (\ref{b3})
yields that $(\nabla\times {\rm curl}u)\cdot n=0$ on the boundary
$\partial\Omega $ (see \cite{bdg}), and $(\ref{a1})_1$ can be
rewritten as \bnn \la{a81}\nabla G=\rho u_t+\rho u\cdot\nabla
u+\mu\nabla\times {\rm curl}u, \enn we have  \bnn \nabla G\cdot
n|_{\partial\Omega}= \rho (u\cdot\nabla) u\cdot n|_{\partial\Omega}
=  -\rho (u\cdot\nabla) n\cdot u|_{\partial\Omega} . \enn Therefore,
(\ref{a81}) yields that $G$ satisfies \bnnn\begin{cases}\Delta
G={\rm div}(\rho u_t+\rho u\cdot\nabla u)\\ \nabla G\cdot
n|_{\partial\Omega}  =  -\rho (u\cdot\nabla) n\cdot
u|_{\partial\Omega}. \end{cases}\ennn Using the $L^q$-estimate for
Neumann problem to the elliptic equation, we have \bnn\la{a82}
\|\nabla G\|_{L^q}\le C\left( \norm[L^q]{\rho u_t} +
\norm[L^q]{u\cdot\nabla u}+\|\rho|u|^2\|_{C(\bar{\Omega})}\right)\le
C(\|\nabla u_t\|_{L^2}+1).\enn We deduce from  (\ref{a50})(resp.
(\ref{a80})),    (\ref{a49})(resp. (\ref{a82})), (\ref{a42}), and
Gronwall's inequality that\be \la{a65}\sup\limits_{0\le t\le
T}\norm[W^{1,q}]{\rho}\le C. \ee  We complete the proof of Lemma
\ref{z3}.

The combination of Lemmas \ref{z2} and  \ref{z3}    is enough to
extend the classical solutions of $(\rho,u)$ beyond $t\ge T^*$. In
fact, in view of (\ref{a42}) and (\ref{a48}), the functions
$(\rho,u)|_{t = T^*} = \lim_{t\rightarrow T^*}(\rho,u)$ satisfy the
conditions imposed on the initial data $(\ref{a10}) $ at the time
$t=T^*.$ Furthermore, \bnnn -\mu\lap u-(\mu + \lambda)\nabla({\rm
div }u) + \nabla P|_{t = T^*} = \lim_{t\rightarrow T^*} (\rho u_t +
\rho u\cdot\nabla u) \triangleq \rho^{\frac{1}{2}} g|_{t = T^*},
\ennn whith $g|_{t = T^*}\in L^2(\O)$. Thus, $(\rho,u)|_{t = T^*} $
satisfies (\ref{a11}) also. Therefore, we can take $(\rho,u)|_{t =
T^*}$ as the initial data and apply the local existence theorem
\cite{K3,K1} to extend our local strong solution beyond $T^*$. This
contradicts the assumption on $T^*$.

\section{Generalization to the heat-conductive flows}
We can modify the previous proof to be fit for the heat-conductive
flows. First, following the proof of Lemma  \ref{z1}  and noticing
that \be\la{d1} P_t + u\cdot\nabla P + \g P\text{div}u =
(\g-1)\kappa\triangle\theta + (\g-1)(\frac{\mu}{2}(\nabla u+\nabla
u^t)^2 + \lambda(\text{div}u)^2), \ee one gets \be\la{d2} \ba
&\frac{d}{dt}\int_{\O }\frac{\mu}{2}|\nabla u|^2 +
\frac{\mu+\lambda}{2}(\text{div}u)^2-P\text{div}udx\\& \quad
+\int_{\O }\rho^{-1}(\mu\triangle u + (\mu +
\lambda)\nabla\text{div}u-\nabla P)^2dx\\ &\le
C(\norm[L^{\infty}]{\mathcal{D}u}+1)\norm[L^2]{\nabla u}^2
+ |\int_{\O}P_t\text{div}udx|\\
& \le C(\norm[L^{\infty}]{\mathcal{D}u}+1)\norm[L^2]{\nabla u}^2 + \ep\norm[H^1]{\nabla u}^2
+ C(\ep)\norm[L^2]{\nabla\theta}^2 \\
& + C(\ep)(1+\norm[L^{\infty}]{\mathcal{D}u} +
\norm[L^{\infty}]{\theta}^2) \norm[L^2]{\nabla u}^2, \quad \forall
0<\ep<1. \ea \ee Multiplying $\theta$ on both sides of the energy
equation, one has after integration by parts that \be\la{d3}
\p_t\int_{\O}\rho\theta^2dx + 2\kappa\int_{\O}|\nabla\theta|^2dx \le
C\norm[L^{\infty}]{\theta}^2 \norm[L^2]{\nabla u}^2. \ee Combining
(\ref{a24}), (\ref{d2}), (\ref{d3}) and applying Gronwall's
inequality, we easily deduce that \be\la{d4} \ba \sup_{0\le t\le
T}(\norm[L^2]{\nabla u}^2 + \norm[L^2]{\nabla\rho}^2 +
\norm[L^2]{\rho^{\frac{1}{2}}\theta}^2 ) + \int_0^T(\norm[
H^1]{\nabla u}^2 + \norm[L^2 ]{\nabla\theta}^2)dt\le C. \ea
\nonumber\ee The higher regularity of $(\rho,u,\theta)$ can be
obtained following the proof of \cite{hl}.

{\bf Acknowledgement}  This research is supported in part by Zheng
Ge Ru Foundation, Hong Kong RGC Earmarked Research Grants
CUHK4028/04P, CUHK4040/06P, CUHK4042/08P, and the RGC Central
Allocation Grant CA05/06.SC01.  The research of J. Li
 is partially supported by    NNSFC No.10601059\&10971215.

\end{document}